\documentstyle[11pt,aaspp4,flushrt,natbib209]{article}
\slugcomment{\it Submitted to the Astrophysical Journal Supplement}
\lefthead{Stern et al.}
\righthead{Cosmic Chronometers. II.  Spectroscopic Catalog}

\def\eg{{e.g.,~}}

\def\deg{\ifmmode {^{\circ}}\else {$^\circ$}\fi}
\def\kms{\ifmmode {\rm\,km\,s^{-1}}\else
    ${\rm\,km\,s^{-1}}$\fi}
\def\ergcm2s{\ifmmode {\rm\,ergs\,cm^{-2}\,s^{-1}}\else
    ${\rm\,ergs\,cm^{-2}\,s^{-1}}$\fi}
\def\ergAcm2s{\ifmmode {\rm\,ergs\,cm^{-2}\,s^{-1}\,\AA^{-1}}\else
    ${\rm\,ergs\,cm^{-2}\,s^{-1}\,\AA^{-1}}$\fi}
\def\ergs{\ifmmode {\rm\,ergs\,s^{-1}}\else
    ${\rm\,ergs\,s^{-1}}$\fi}
\def\kmsMpc{\ifmmode {\rm\,km\,s^{-1}\,Mpc^{-1}}\else
    ${\rm\,km\,s^{-1}\,Mpc^{-1}}$\fi}

\def\spose#1{\hbox to 0pt{#1\hss}}
\def\simlt{\mathrel{\spose{\lower 3pt\hbox{$\mathchar"218$}}
     \raise 2.0pt\hbox{$\mathchar"13C$}}}
\def\simgt{\mathrel{\spose{\lower 3pt\hbox{$\mathchar"218$}}
     \raise 2.0pt\hbox{$\mathchar"13E$}}}

\def\plotone#1{\centering \leavevmode
\epsfxsize=\columnwidth \epsfbox{#1}}

\def\plotfiddle#1#2#3#4#5#6#7{\centering \leavevmode
\vbox to#2{\rule{0pt}{#2}}
\includegraphics{#1}}

\begin{document}

\title{Cosmic Chronometers:  Constraining the Equation of State of Dark
Energy.  II.  A Spectroscopic Catalog of Red Galaxies in Galaxy Clusters}

\author{Daniel Stern\altaffilmark{1},
Raul Jimenez\altaffilmark{2},
Licia Verde\altaffilmark{2}, 
S. Adam Stanford\altaffilmark{3},
Marc Kamionkowski\altaffilmark{4}}

\altaffiltext{1}{Jet Propulsion Laboratory, California Institute of
Technology, 4800 Oak Grove Drive, Mail Stop 169-506, Pasadena, CA
91109 [e-mail: {\tt stern@zwolfkinder.jpl.nasa.gov}]}

\altaffiltext{2}{ICREA \& Institute of Sciences of the Cosmos (ICC),
University of Barcelona, Barcelona 08028, Spain}

\altaffiltext{3}{University of California, Davis, CA 95616 and
Institute for Geophysics and Planetary Physics, Lawrence Livermore
National Laboratory, Livermore, CA 94551}

\altaffiltext{4}{California Institute of Technology, Mail Stop 350-17,
Pasadena, CA 91125}

\begin{abstract} 

We present a spectroscopic catalog of (mostly) red galaxies in 24
galaxy clusters in the redshift range $0.17 < z < 0.92$ obtained
with the LRIS instrument on the Keck~I telescope.  Here we describe
the observations and the galaxy spectra, including the discovery
of three cD galaxies with LINER emission spectra, and the spectroscopic
discovery of four new galaxy-galaxy lenses in cluster environments.

\end{abstract}


\section{Introduction}

The nature of the physics driving cosmic acceleration is perhaps
the biggest question facing physics today.  Huge resources and large
collaborations are now being amassed to determine the dark-energy
equation-of-state parameter $w\equiv p/\rho$, relating the cosmic
pressure $p$ to the energy density $\rho$ (in units of $c \equiv
1$).  The value of $w$ could either be constant, as in the case
of a cosmological constant ($w = -1$), or time-dependent, as in
the case of a rolling scale field or ``quintessence'' \citep{Peebles:88,
Caldwell:98}.  Any such behavior would have far-reaching implications
for particle physics.  The avenues that are now receiving the most
attention are supernova searches, weak lensing, baryon acoustic
oscillations, and cluster counts.  However, none of these will be
free from systematics, and it is still not clear which is the most
promising approach.  Most likely, multiple approaches will be
required, and new ideas still need to be explored.

A new approach, proposed by \citet{Jimenez:02}, is to measure the
relative ages of luminous red galaxies, a probe that is particularly
sensitive to the {\em variation} of $w(z)$ with redshift $z$.  This
method was demonstrated to work in \citet{Jimenez:03} with a set
of low-redshift ($z \simlt 0.25$) Sloan Digital Sky Survey (SDSS)
luminous red galaxies from the sample of \citet{Eisenstein:01}
supplemented with approximately two dozen moderate-redshift ($z
\simlt 1$) early-type galaxies observed with Keck.  That work found
$w \leq -0.8$ at the 68\% confidence level.  Using the SDSS sample
to measure $dz/dt$ at $z \sim 0$, that work also derived an independent
estimate for the Hubble constant, $H_0 = 69 \pm 12 \kmsMpc$.  Although
this new cosmological test faces challenges from astrophysical
uncertainties, these are not necessarily any more daunting than
those associated with the more classical dark energy probes.

We describe here the results of a several night experiment with the
Keck~I telescope to obtain spectra of early-type galaxies in clusters
at $0.2 \simlt z \simlt 1$.  As demonstrated in \citet{Treu:05},
the most massive early-type galaxies have the highest formation
redshifts and are thus best suited to this experiment.  Such galaxies
are concentrated in galaxy clusters, with the additional benefit
that {\it GALEX} observations show that cluster elliptical galaxies
have less UV emission than field elliptical galaxies
\citep[e.g.,][]{Schawinski:07}.  This implies that cluster ellipticals
have the lower levels of ``frosting'' by recent star formation and
thus are best suited for this experiment.

The multiplexing and unparalleled blue sensitivity of the Low
Resolution Imaging Spectrometer \citep[LRIS;][]{Oke:95} on the
Keck~I telescope is a crucial aspect of this project, allowing
simultaneous and accurate spectrophotometry of up to 20 cluster
galaxies over a very wide spectral range, probing down to 2000 \AA\
in the restframe.  The importance of the UV stellar breaks at
restframe 2640~\AA\ and 2900~\AA\ (B2640, B2900) was shown by
\citet{Fanelli:92} and were used by \citet{Dunlop:96} and
\citet{Spinrad:97} to determine the age of the high-redshift,
quiescent radio galaxy LBDS~53W091.  A comprehensive study by
\citet{Dorman:03} dramatically shows that the restframe UV spectral
region can be modeled accurately and, more importantly, that UV
light (bluewards of the 4000~\AA\ region) can help to break the
age-metallicity degeneracy.

\citet[][Paper~I]{Stern:09a} presents the cosmological results from
this experiment and the derived constraints on both the equation
of state of dark energy and the value of the Hubble constant, $H_0$.
This paper (Paper~II) presents the data in some detail, including
target selection, observing strategy, data processing (\S 2), and
a catalog of approximately 500 sources in the fields of galaxy
clusters out to $z \approx 1$ (\S 3.1).  It is hoped that this
resource will be useful for a number of other projects, such as
more classic studies of galaxy and galaxy cluster evolution.  In
particular, we identified several interesting cD galaxies (\S 3.2)
and we serendipitously found four new galaxy lenses (\S 3.3).  Since
most of the target sample are well-known clusters, some of these
lenses already have {\it Hubble Space Telescope} imaging, and a
more detailed analysis of the lens systems will be presented in
Moustakas et al. (in prep.).  During the course of this program,
we also observed and derived the mass of a stellar black hole in
the Virgo globular cluster RZ~2109 \citep{Zepf:08}.

\section{Data}

Nine nights were awarded for this experiment between February 2007
and September 2008 (Table~\ref{table.obsrun}).  Unfortunately, five
of the nights were completely lost to weather (four nights) or
instrument problems (one night).  In particular, our two night
observing run in August 2007 was lost to a combination of Hurricane
Flossie (category four), two earthquakes (magnitude 5.4 and 4.0),
and a tsunami warning (due to a magnitude 8.0 earthquake in Peru).
Of the four dedicated nights during which we obtained data, two (in
December 2007) suffered from both a bright ($> 50\%$) moon and poor
conditions (thick cirrus and $\simgt 1\farcs5$ seeing).  As shown
in Table~\ref{table.obsrun}, we obtained a handful of additional
observations during nights dedicated to other programs.

\subsection{Target Selection}

We targeted rich galaxy clusters in order to obtain an as large as
possible sample of red galaxies over the redshift range $0.2 < z <
1$.  Most of the clusters are well-known, rich X-ray clusters from
a variety of samples such as the \citet{Abell:58} catalog, the {\it
ROSAT} Cluster Survey \citep[RCS;][]{Rosati:98} and the Massive
Cluster Survey \citep[MACS;][]{Ebeling:01}.  In the redshift range
$0.5 < z < 1$, fewer rich X-ray clusters are known, so we targeted
(and confirmed) two of the richest {\it Spitzer} mid-infrared
selected cluster candidates from the IRAC Shallow Cluster Survey
\citep{Eisenhardt:08}.  Many of the targeted clusters were also
observed in the near-infrared cluster survey of \citet{Stanford:02},
which provided a valuable and consistent astrometric resource for
slitmask designs.  The list of clusters targeted is presented in
Table~\ref{table.clusters}.

The galaxy cluster sample was chosen to meet the competing demands
of providing a good distribution of cluster redshifts for the
cosmological experiment, a good right ascension distribution for
the assigned nights, as well as a good right ascension distribution
of brighter clusters to observe during poor conditions.  Once a
cluster was chosen for observation, selecting early-type galaxy
cluster members with which to fill the slitmasks was an intensive
process.  We relied heavily on the NASA/IPAC Extragalactic Database
(NED) to identify cluster members in the literature.  For many
clusters, we were able to spectroscopically target known cluster
members that had no known signatures of star formation or AGN
activity.  Some clusters also had published lists of candidate
cluster members based on either morphology or red colors from the
optical to the near-infrared \citep[\eg][]{Stanford:02}.  For the
Bo\"otes clusters in the IRAC Shallow Survey \citep{Eisenhardt:04,
Ashby:09}, photometric redshifts based on optical thru mid-infrared
data were used to select candidate early-type cluster members
\citep{Brodwin:06}.  Many of the clusters also had publicly available
images in the {\it Hubble Space Telescope} archive from which we
were able to morphologically and/or color select candidate cluster
members.

One challenge of the myriad source lists used to populate the masks
was that each source list was based on a slightly different astrometric
reference frame.  For the mask design, we required consistent
astrometry for all candidates across the full $5\arcmin \times
7\arcmin$ LRIS field, including a minimum of three alignment stars
brighter than $B \sim 20$.  For the lower redshift clusters, publicly
available images from the Palomar Sky Survey or SDSS proved sufficient.
For the higher redshift clusters, however, candidate cluster members
were often too faint to provide robust centroiding in those shallow
data.  For most of these clusters we obtained images with
the roboticized Palomar 60\arcsec\ (P60) telescope \citep{Cenko:06}
which is equiped with a SITe 2048 $\times$ 2048 pixel CCD with a
pixel scale of 0\farcs378 pixel$^{-1}$ and a 12\farcm9 $\times$
12\farcm9 field of view.  For each cluster, we observed two bands 
chosen to straddle the 4000~\AA\ break (D4000) at the cluster
redshift.  These multi-band images, either from the P60 or the
public data sets, provided an additional sample of red
sources near the cluster with which to populate the masks.  Note
that due to the geometric contraints from the mask designs, masks
inevitably including a few targets which were unlikely to be cluster
members.

\subsection{Observing Strategy}

The key goal of this program was to provide high signal-to-noise
ratio, wide wavelength coverage spectroscopy of a large number of
early-type galaxies at moderate redshifts.  These spectra were then
modeled to derive the ages of the galaxy stellar populations.  Since
restframe UV light probes the youngest, most massive stars in a
galaxy, blue sensitivity is crucial for this experiment.  Of all
the optical spectrographs on 8 - 10~m class telescopes currently
available, LRIS on the Keck~I telescope is unique in being the only
dual-beam spectrograph, thus providing sensitive observations across
the entire optical window ($\lambda \sim 3200~{\rm \AA} - 1~\mu$m).

LRIS provides spectra of approximately 25 sources simultaneously
across a $\sim 5\arcmin \times 7\arcmin$ field of view, with a
dichroic splitting the light between the blue and red channels.
For most of the observations, we observed the same mask with two
different dichroics, thereby ensuring that the final spectra have
no gaps which might compromise the stellar population modeling.
Only the three masks observed in February 2005 were observed with
a single dichroic.  LRIS has an ample set of dichroics to choose
from; we consistently selected dichroics that avoided the rest-frame
4000~\AA\ spectral region which includes import stellar population
diagnostics (e.g., [\ion{O}{2}]~$\lambda 3727$, D4000 and the Balmer
break).

We obtained the blue channel data with the 400 lines mm$^{-1}$
grism, which has a central wavelength of 3400~\AA\ and a spectral
range of  4450~\AA.  We obtained the red channel data with the 400
lines mm$^{-1}$ grating, which has a central wavelength of 8500~\AA\
and a spectral range of 3800~\AA.  Combining the blue and red channel
data from the two dichroic settings, sources typically had final
spectra which spanned the entire $\sim 3200$~\AA\ $- 1~\mu$m optical
window, albeit with higher noise at the short and long wavelength
extremes.  Based on analysis of sky lines, sources filling the
1\farcs5 wide slitlets used for these observations have resolution
$\lambda / \Delta \lambda \sim 500$ and $\sim 650$ for the blue and
red channels, respectively.  Standard stars from \citet{Massey:90}
were observed with the same instrument configuration for the purposes
of spectrophotometric calibration.

Observations were generally obtained with two dithered exposures
per dichroic configuration with typical integration times of 900~s
to 1800~s, depending on the cluster redshift and observing conditions.
This allowed both improved cosmic ray rejection and, by pair-wise
subtraction of the red images, removal of the fringing which strongly
affects the long wavelength ($\lambda \simgt 7200$ \AA) LRIS data.
This required minimum slitlet lengths of approximately 10\arcsec.
Since LRIS has an atmospheric dispersion corrector, mask position
angles were optimized based on the cluster orientation and no special
attention was necessary to align the masks with the parallactic
angle.

\subsection{Reductions}

We processed the spectroscopic data with {\tt BOGUS}\footnote{Available
upon request from the first author.}, which is an IRAF routine
designed for two-dimensional processing of multislit data and was
written by D.~Stern, A.~Bunker, and S.~A.~Stanford.  After gain and
overscan correction of the raw two-dimensional images, {\tt BOGUS}
basically splits the mask into individual slitlets and processes
each slitlet using standard optical longslit techniques.  After
flattening the spectrum with either domeflats (recommended for the
red channel of LRIS) or twilight flats (recommended for the blue
channel of LRIS), cosmic rays are identified from unsharp masking
of the images, sky lines are subtracted using a low order polynomial
fit to each column, and images are shifted by integer pixels in the
spatial and dispersion directions and recombined.  For the red
channel data, an additional step of pair-wise image subtraction
improves fringe subtraction at long wavelength.  As a final step,
{\tt BOGUS} shifts each of the slitlets to roughly align them in
the wavelength direction.  This both simplifies wavelength calibration
and the rapid visual identification of spectroscopically-confirmed
cluster members.

After the two-dimensional processing provided by {\tt BOGUS}, we
extracted the spectra using 1\farcs5 wide extraction traces using
the {\tt APALL} procedure within IRAF.  We extracted arc lamps in
an identical manner and used them to do a first pass wavelength
calibration of the data.  This typically relied on a fourth order
polynomial wavelength solution, providing a $\approx 0.5$~\AA\ RMS
to the blue channel of LRIS and a $\approx 0.1$~\AA\ RMS to the red
channel of LRIS.  As a final step in the wavelength calibration,
these lines were linearly shifted based on the sky lines and we
conservatively estimate that the wavelength solutions are robust
to better than 1~\AA.  We flux calibrated the spectra using
spectrophotometic standards observed during each observing run.

At this point, each slitmask target generally has four spectra: the
blue and red channel observations for each of the two dichroics.
To combine the spectra into a single, final spectrum for the stellar
population synthesis analysis, we did the following.  Spectra were
trimmed at their blue and red ends to restrict coverage to regions
of more robust spectrophotometry.  Generally, wavelengths blueward
of observed $\sim 3500$~\AA\ were removed due to their lower
signal-to-noise.  Pixels within 50~\AA\ of the dichroics were
eliminated.  At long wavelength, spectral trimming depended on
individual analysis of the spectra.  Fainter targets and/or targets
observed with no dithering often showed significant systematic
glitches in their long wavelength data, and thus were trimmed at
relatively blue wavelengths ($\sim 7500$~\AA).  Other sources had
robust spectra out to $\sim 9500$~\AA.  Since the observations were
not all obtained in photometric conditions, the final combination
of the spectra required multiplicative scaling of their calibrated
spectra, which generally was normalized to the blue channel data
with the longer wavelength dichroic.  We derived statistical error
spectra assuming Poisson uncertainties of sky plus science target
counts within the extraction regions.  For observations where an
extra step of fringe subtraction was applied, the error spectra
were increased by $\sqrt{2}$ to account for the statistical noise
hit from that procedure.  During the final step of combining the
multiple spectra (which generally were all obtained with the same
exposure time), the error spectra were averaged and scaled down by
$\sqrt{2}$.

%

\section{Results}

\subsection{Spectra of Galaxies in Clusters}

\begin{figure}
\plotfiddle{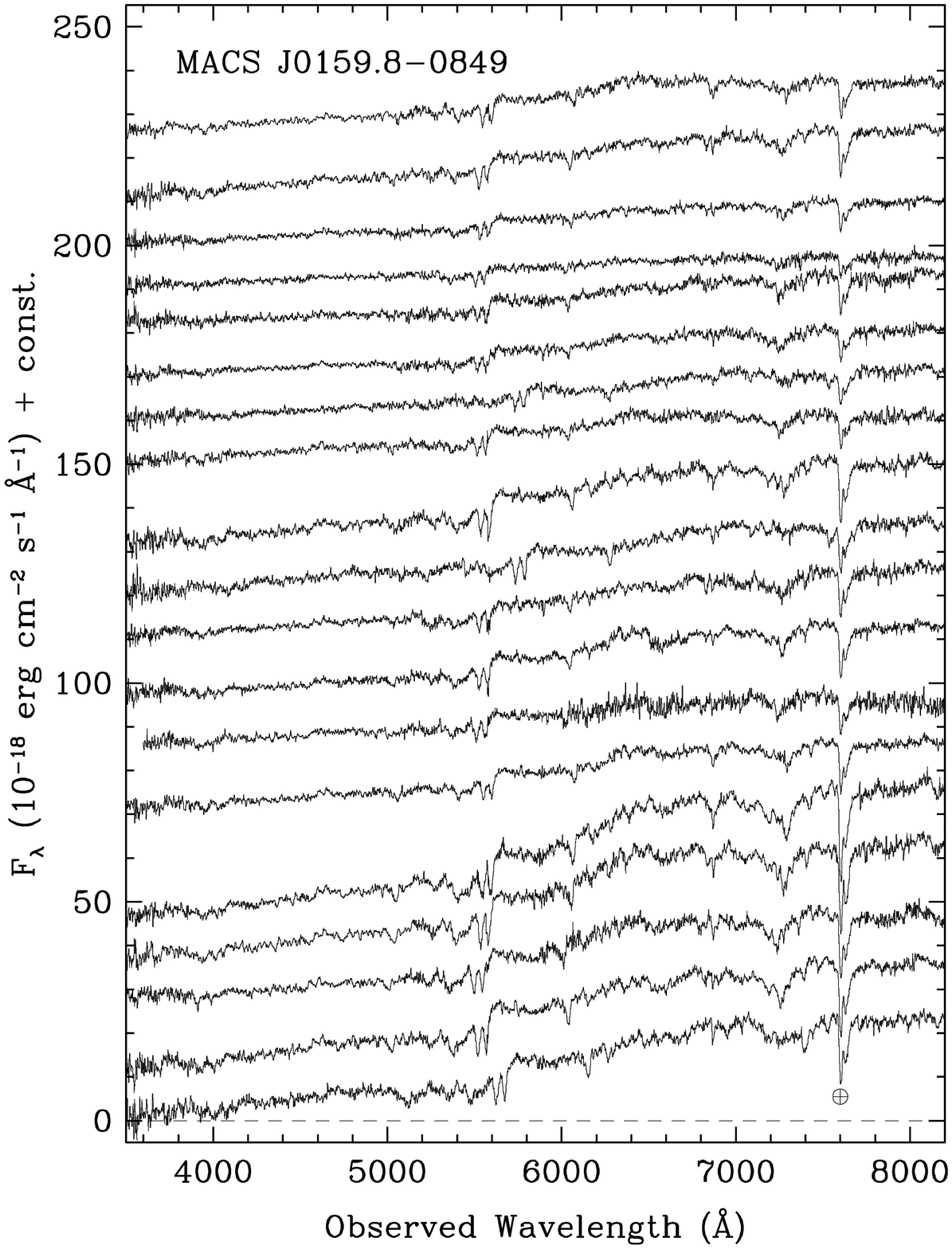}{6.5in}{0}{70}{70}{-220}{-20}
\epsscale{.80}
\caption{Results of a typical mask, MACS~J0159.8$-$0849 at $z=0.405$, observed
on UT 2008 September 3.  Telluric A-band absorption is indicated.
\label{fig:macs0159}}
\end{figure}

\begin{figure}
\plotfiddle{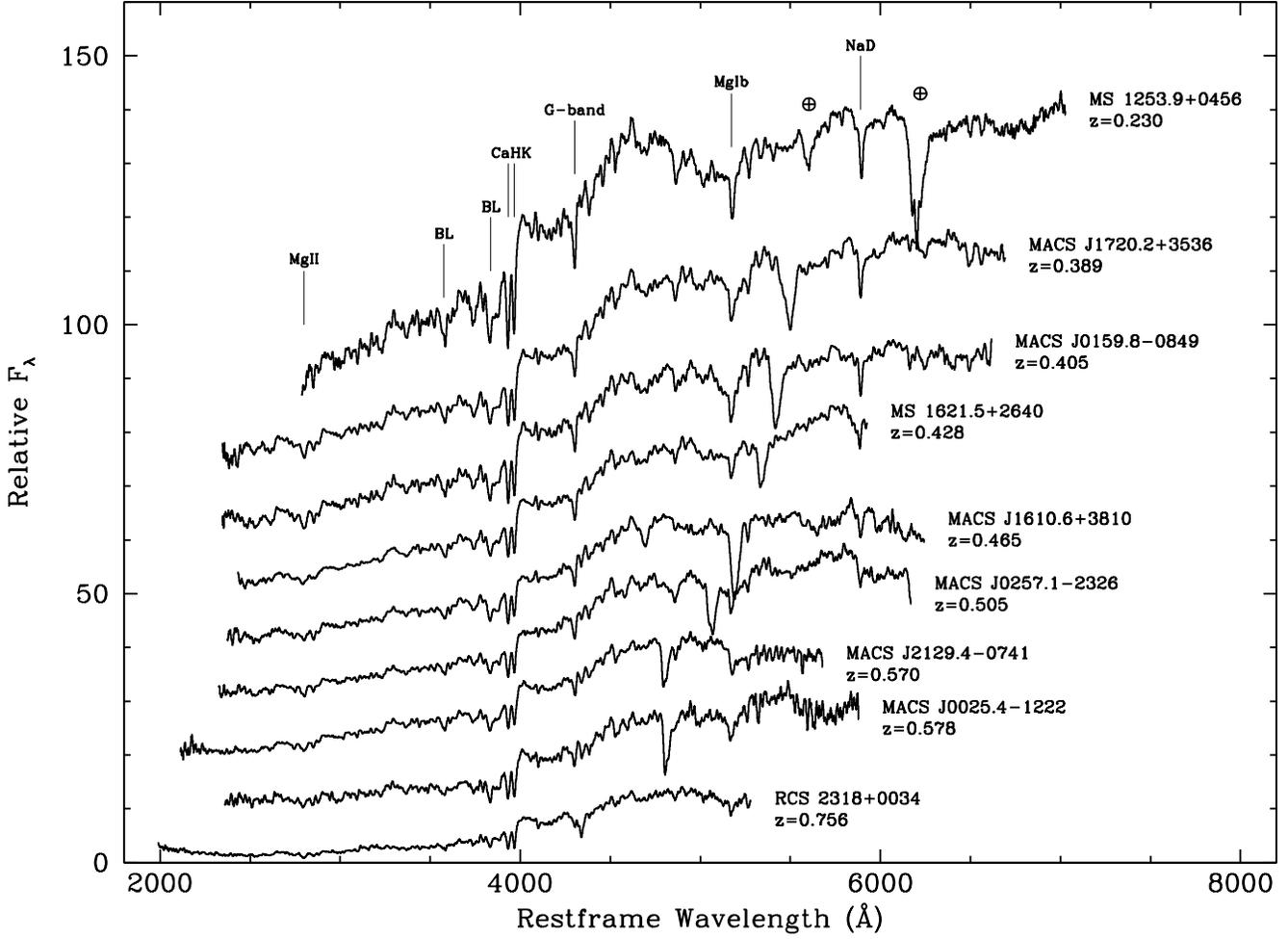}{5.5in}{-90}{70}{70}{-280}{420}
\epsscale{.80}
\caption{Averaged spectra of cluster ellipticals from the observing
runs in July and September 2008, plotted as a function of redshift.
Primary spectral features are indicated.  Telluric A-band (7600
\AA) and B-band (6880 \AA) absorption are indicated for the lowest
redshift cluster (top); these features shift to shorter restframe
wavelengths for the higher redshift clusters.
\label{fig:atlas}}
\end{figure}

Figure~\ref{fig:macs0159} presents the results from the a typical
mask:  MACS~J0159.8$-$0849 at $z = 0.405$, observed on UT 2008
September 3.  Nearly 20 early-type cluster members were obtained
simultaneously, each clearly showing an evolved stellar population
with strong Ca~H and K absorption and a prominent D4000 break.
Figure~\ref{fig:atlas} averages spectra of cluster members for masks
observed in late 2008 and shows a clear sequence with redshift:
clusters at lower redshift contain older stellar populations,
recognized from their redder spectral energy distributions (SEDs)
and larger D4000 breaks.

Table~\ref{table.spectra} presents the results from the spectroscopy.
The spectra are classified both on the basis of their quality and
their spectral class.  Quality ``A'' indicates very reliable
redshifts, generally based on multiple, well-observed features.
Quality ``B'' indicates reliable redshifts, but often based on a
single feature --- e.g., a single emission line is observed and is
assumed to be either Ly$\alpha$ or [\ion{O}{2}], or a break is
identified, but not with sufficient fidelity to derive a precision
redshift.  Such redshifts are likely correct, though an occasional
misidentification is possible.  Based on a comparison of redshifts
derived from multiple features, quality ``A'' redshifts are
conservatively estimated to have uncertainties of 0.002 in redshift.
Assuming the features are correctly identified, quality ``B''
emission line redshifts will have comparable uncertainties, while
quality ``B'' absorption line redshifts will have uncertainties
twice as large.  Our internal processing also included quality ``C''
(tentative) and ``F'' (uncertain) flags; such sources are omitted
from Table~\ref{table.spectra}.  

We assign each of the spectra one of four spectroscopic classes (or
a hybrid of these classes).  Spectroscopic class ``E'' refers to
early-type or elliptical galaxy spectra, showing only spectral
breaks and absorption lines.  Such sources were the primary goal
of this program.  Spectral class ``S'' refers to late-type or spiral
galaxy spectra, showing emission lines such as [\ion{O}{2}], H$\beta$,
[\ion{O}{3}] and H$\alpha$, typical of star forming galaxies.  We
identified several stars during this program, tabulated as spectral
class ``$\star$''.  Many are late-type stars of spectral class M,
mistakenly targeted on the basis of their red colors mimicing that
of moderate-redshift early-type galaxies.  A handful of early-type
stars were also serendipitously observed.  Finally, we observed a
handful of active galaxies, listed as spectral class ``AGN.''

\begin{figure}
\plotone{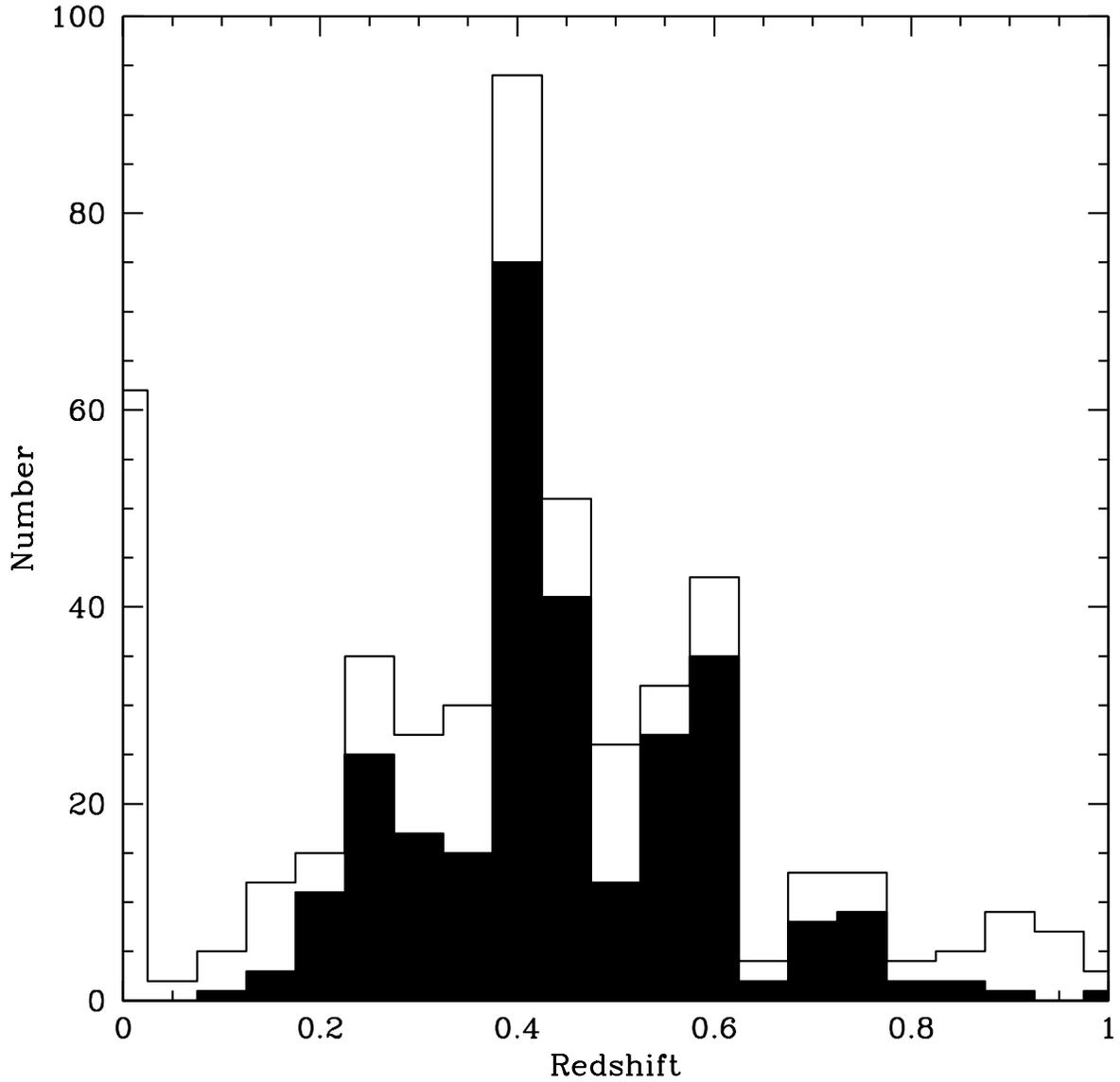}
\caption{Histogram of redshifts derived from this program.  The
open histogram shows all 496 sources, while the solid histogram
only shows the highest quality (quality ``A'') absorption-line (class
``E'') redshifts.  Note the high fraction of early-type galaxies from
this program, much higher than would be found in a field galaxy
survey.
\label{fig:zhist}}
\end{figure}

In total, we obtained 496 redshifts in galaxy cluster fields during
this program, of which the vast majority (464, or 94\%) are quality
``A'' (Figure~\ref{fig:zhist}).  For the targeted cluster early-type
galaxy sample, we obtained a total of 278 sources, of which 260 are
quality ``A.'' Note that just because a source is of quality ``A''
doesn't ensure that it will be useful for the cosmic chronometer
experiment.  A target might have a reliable redshift due to the
detection of specific features, while the continuum might be of low
signal-to-noise ratio due to poor fringe subtraction, contamination
from a nearby source, or simply due to the observing conditions.
\citet{Stern:09a} analyzes this sample, deriving stellar population
ages by modeling the spectra.


\subsection{cD Galaxies}

\begin{figure}
\plotfiddle{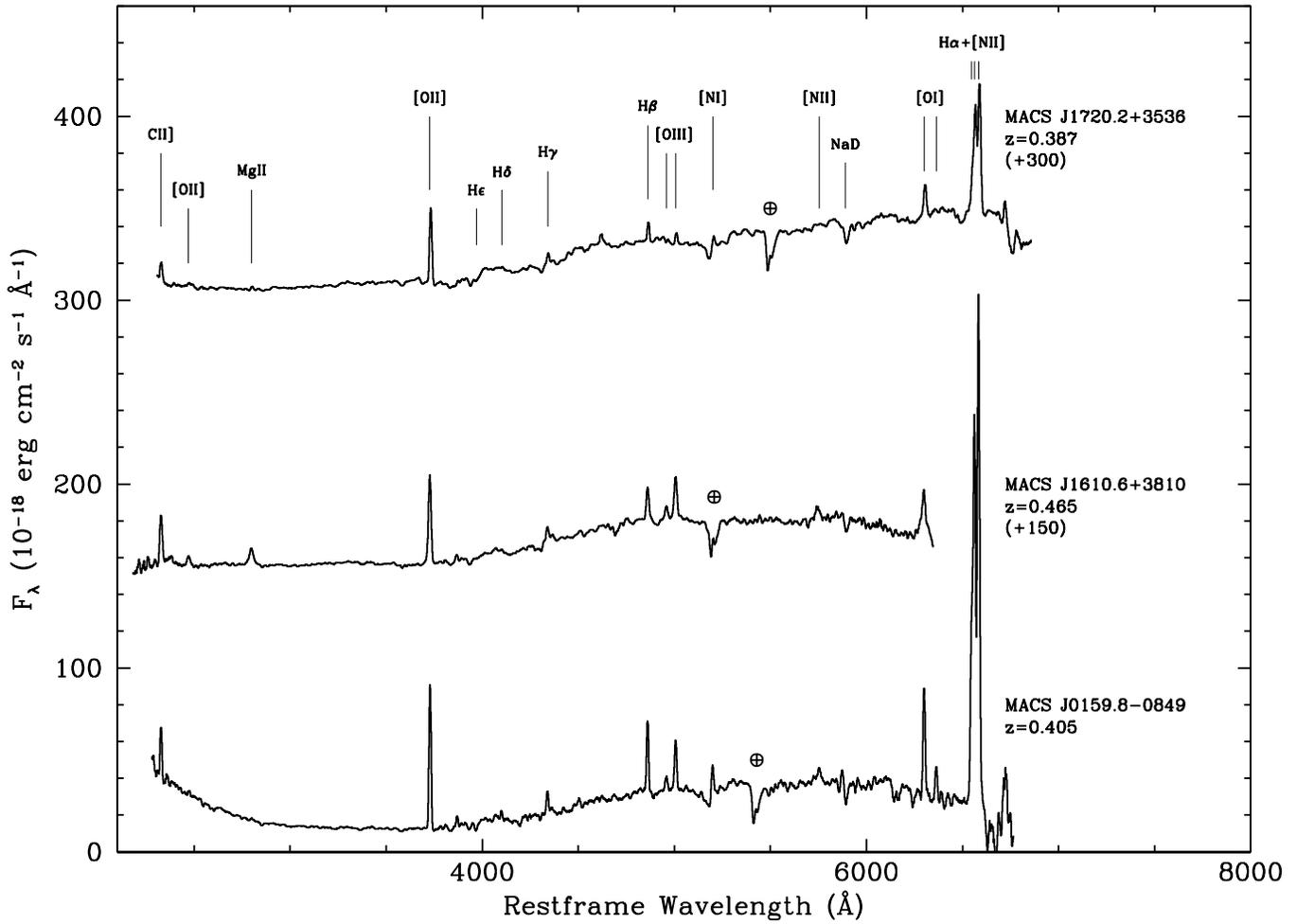}{5.5in}{-90}{70}{70}{-280}{420}
\epsscale{.80}
\caption{Spectra of three cluster cD galaxies showing strong emission
lines, typical of LINERs.
\label{fig:cD}}
\end{figure}

Three of the central cD galaxies observed during this program have
interesting spectra showing multiple emisison lines superposed on
a red, evolved stellar population (Figure~\ref{fig:cD}).  The spectra
are similar to that of NGC~1275, the central galaxy in the Perseus
cluster \citep[e.g.,][]{Sabra:00}, and are clearly classified as
LINER-like spectra.  Indeed, $\log$ ([\ion{O}{1}]~6300/[\ion{O}{3}]~5007)
ranges from 0.13 to 0.59 for the three galaxies, which is much
larger than typical of star-forming galaxies in the SDSS
\citep{Brinchmann:08}.  All three galaxies are detected by the Faint
Images of the Radio Sky at Twenty cm survey \citep[FIRST;][]{Becker:95},
with 1.4~GHz flux densities of 31.43~mJy, 5.59~mJy and 16.75~mJy
for MACS~J015949.3$-$84958, MACS~J162124.7+381008 and
MACS~J172016.7+353626, respectively.

While typically only a modest fraction ($\approx 15\%$) of brightest
cluster galaxies (BCGs) show strong optical line emission, optical
line emission is quite common ($71^{+9}_{-14} \%$) for BGCs in
cooling flow clusters \citep{Edwards:07}.  The line emission is
presumed related to the cooling of X-ray gas at the cluster center.
For the well-studied, local example of NGC~1275, the line emission
is concentrated in a spectacular network of filaments extending
over several arcminutes.  The filaments are thought due to compressed,
cooling intracluster gas within a relativistic plasma ejected by
the active nucleus of NGC~1275 \citep{Conselice:01}.  NGC~1275 also
shows evidence of recent star formation.  Understanding such systems
in detail is likely important for probing the physics of feedback in
massive galaxies.

\subsection{Lensed Galaxies}

\begin{figure}
\plotone{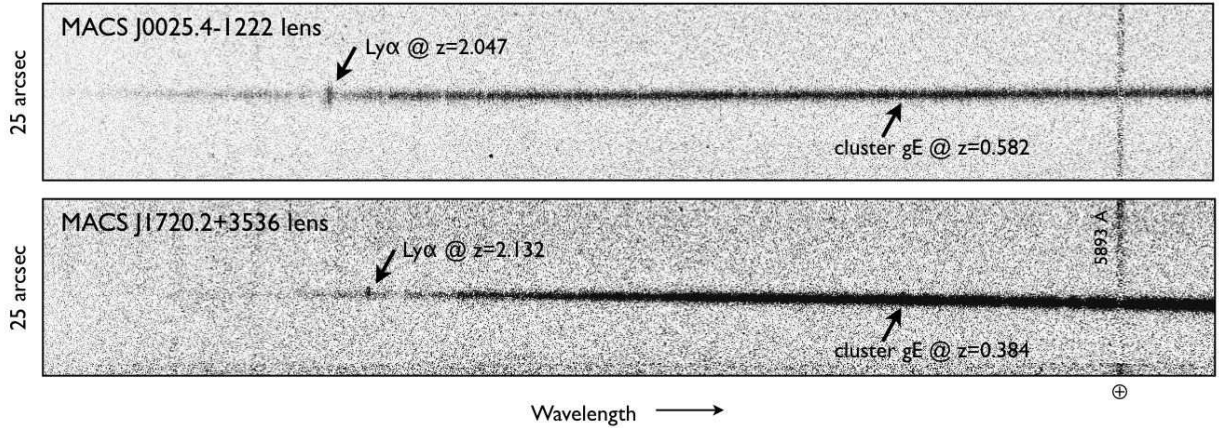}
\caption{Two-dimensional processed spectra of two of the newly
identified galaxy-galaxy lenses in cluster environments.  Both
images, obtained with the D680 grating and the blue channel of LRIS,
span approximately from 2900~\AA\ to 6150~\AA\ in the dispersion
(horizontal) axis and 25\arcsec\ in the spatial (vertical) axis.
Residuals from telluric NaD~5893~\AA\ emission are visible at the
long wavelength ends of the spectra.  Both spectra show strong,
extended Ly$\alpha$ emission superposed atop red emission from
early-type members of the targeted MACS clusters.
\label{fig:lens2D}}
\end{figure}

\begin{figure}
\plotfiddle{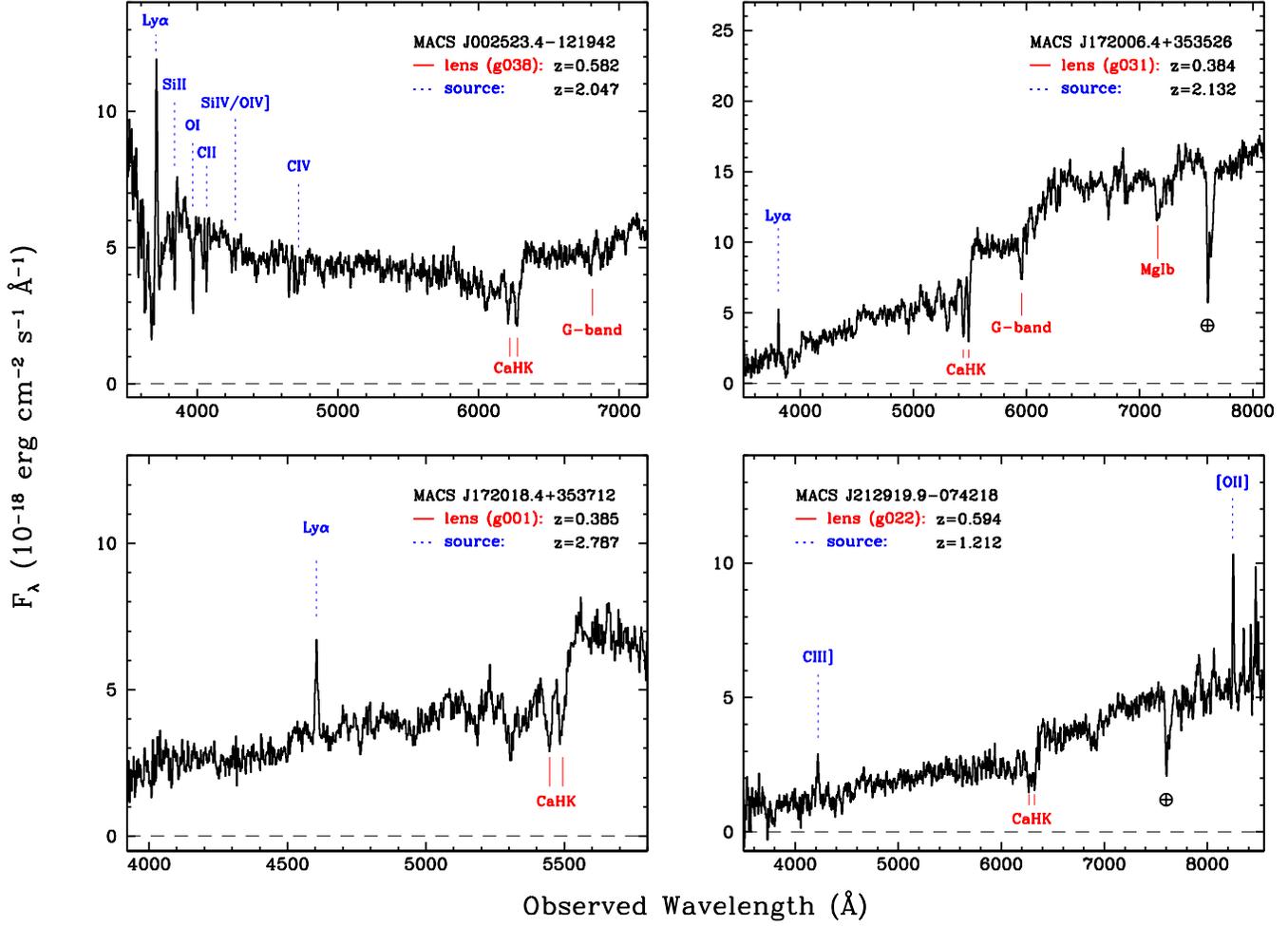}{4.0in}{-90}{70}{70}{-280}{420}
\caption{Extracted spectra of the four newly identified galaxy-galaxy
lenses in cluster environments.
\label{fig:lens1D}}
\end{figure}

Four of the cluster galaxies observed as part of this program show
additional emission lines at blue wavelengths superposed on cluster
early-type spectra.  These emission lines are spatially extended
in three of the four sources (Figure~\ref{fig:lens2D}).  In all
four sources the emission lines do not match the absorption line
redshift (Figure~\ref{fig:lens1D}).  The spectra are reminiscent
of lenses spectroscopically identified in the SDSS luminous red
galaxy sample, such as the $z=2.7$ Einstein cross identified by
\citet{Bolton:06} and lower redshift strong lenses identified by
the Sloan Lens ACS (SLACS) survey \citep{Bolton:06b}.  Two of the
four new lenses have publicly available, two-band Advanced Camera
for Surveys (ACS) images in the {\it Hubble} Legacy Archive.  These
images, presented in Figure~\ref{fig:lensimage}, clearly show
strongly lensed background galaxies behind red, early-type galaxies.
Such systems are valuable probes of the lensing galaxy mass and
mass profile.  Moustakas et al. (in prep.) presents a detailed
analysis of these new systems.

\section{Summary}

We present a catalog of nearly 500 redshifts obtained in the fields
of 24 galaxy clusters, including nearly 300 early-type cluster
members.  Paper~I derives ages for this sample and uses their ages
to probe cosmological parameters using the differential age or
cosmic chronometer test.  This database will also be useful for
studying the properties of galaxy clusters, modeling their mass
distributions, and understanding the formation mechanism of both
clusters and the constituent galaxy populations.  We identify three
interesting central galaxies showing strong, LINER-like spectra,
typical of cooling flow clusters.  We also serendipitously identify
four new galaxy-galaxy lenses on the outskirts of galaxy clusters.

\acknowledgements 

\begin{figure}
\plotone{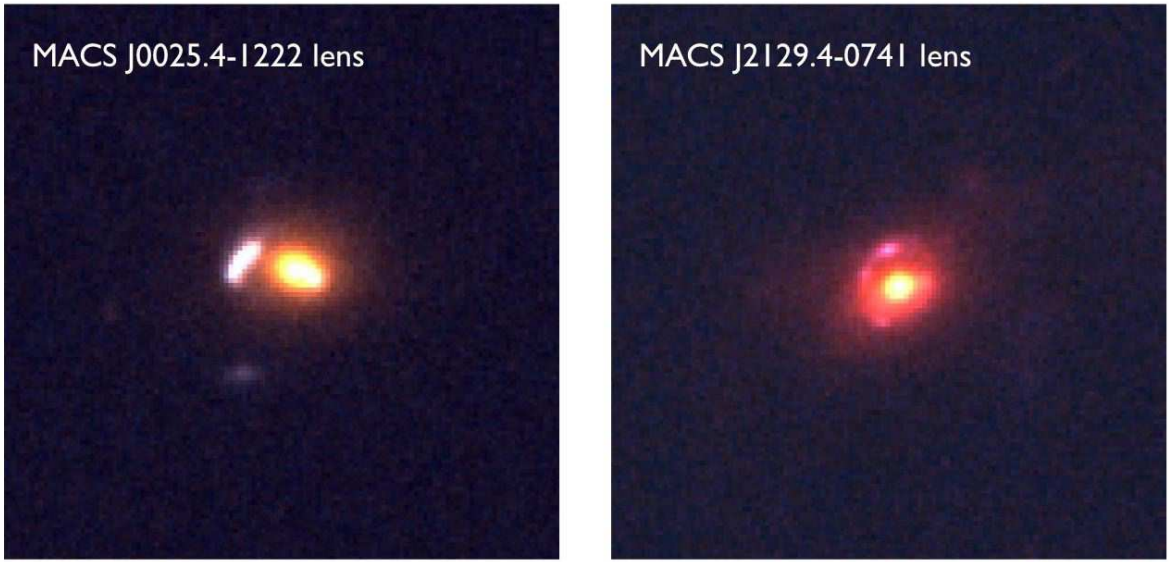}
\caption{{\it HST}/ACS images of two of the newly identified
galaxy-galaxy lenses in cluster environments.  Both images, obtained
from the {\it Hubble} Legacy Archive, are false color images created
from F555W and F814W observations.  Images are approximately 6\arcsec\
on a side, with North up and East to the left.
\label{fig:lensimage}}
\end{figure}

The authors wish to recognize and acknowledge the very significant
cultural role and reverence that the summit of Mauna Kea has always
had within the indigenous Hawaiian community; we are most fortunate
to have the opportunity to conduct observations from this mountain.
We thank M.~Kasliwal for assistance with the P60 scheduling, and
F.~Harrison and R.~Griffith for assisting with the March 2009 Keck
observations.  We also thank A.~Barth for an interesting discussion
of the cD galaxy spectra.  The work of DS was carried out at Jet Propulsion
Laboratory, California Institute of Technology, under a contract
with NASA.  RJ and LV acknowledge support from the Spanish Ministerio
de Ciencia e Innovacion and the European Union FP7 program.  MK was
supported by DoE DE-FG03-92-ER40701 and the Gordon and Betty Moore
Foundation.  This research made use of the NASA/IPAC Extragalactic
Database (NED) which is operated by the Jet Propulsion Laboratory,
California Institute of Technology, under contract with NASA.
Figure~\ref{fig:lensimage} is based on observations made with the
NASA/ESA {\it Hubble Space Telescope}, and obtained from the {\it
Hubble} Legacy Archive, which is a collaboration between the Space
Telescope Science Institute (STScI/NASA), the Space Telescope
European Coordinating Facility (ST-ECF/ESA) and the Canadian Astronomy
Data Centre (CADC/NRC/CSA).

\eject

\scriptsize
\begin{deluxetable}{lccl}
\tablecaption{Cosmic Chronometer Observing Runs.}
\tablehead{
\colhead{UT Date} &
\colhead{\% Dark} &
\colhead{PI} &
\colhead{Conditions/Comments}}
\startdata
2005 Feb 10            & 94    & Stanford & \\
2007 Feb 17-18$^\ddag$ & 100   & Kamionkowski & weathered out \\
2007 Aug 15-16$^\ddag$ & 92-87 & Kamionkowski & Hurricane Flossie, two earthquakes, tsunami warning \\
2007 Dec 17-18$^\ddag$ & 48-40 & Kamionkowski & poor conditions and bright moon \\
2008 Jul 1$^\ddag$     & 89    & Kamionkowski & \\
2008 Sep 2-3$^\ddag$   & 93-87 & Kamionkowski & 1st night lost to mechanical problems \\
2009 Mar 2-3           & 62-52 & Harrison     & poor conditions and bright moon \\
\enddata

\tablecomments{\ddag: Nights dedicated to cosmic chronometer
experiment.  Note the poor track record for the dedicated nights
with only two successful nights out of nine allocations, thus
traumatizing the PI during his first foray into observational
astrophysics.  Other observations were generally just a mask or two
observed during nights focused on other programs.}

\label{table.obsrun}
\end{deluxetable}
\normalsize

\scriptsize
\begin{deluxetable}{cccclll}
\tablecaption{Galaxy Clusters Observed.}
\tablehead{
\colhead{Galaxy Cluster} &
\colhead{RA} &
\colhead{Dec} &
\colhead{Redshift} &
\colhead{UT Date} &
\colhead{PA (deg)} &
\colhead{Comments}}
\startdata
MS~0906.5+1110      & 09:09:12.7 &   +10:58:29 & 0.172 &  2005 Feb 10    &  $-$70.1 & \\
MS~1253.9+0456      & 12:56:00.0 &   +04:40:00 & 0.230 &  2008 Jul 1     &     42.7 & \\
Abell 1525          & 12:21:57.8 & $-$01:08:03 & 0.260 &  2007 Dec 18    &    122.6 & \\
MS~1008.1$-$1224    & 10:10:32.3 & $-$12:39:52 & 0.301 &  2007 Dec 17    &    115.5 & \\
CL~2244$-$0205      & 22:47:13.1 & $-$02:05:39 & 0.330 &  2007 Dec 17-18 & $-$124.4 & poor conditions \\
Abell 370           & 02:39:53.8 & $-$01:34:24 & 0.374 &  2007 Dec 18    &     70.0 & \\
MACS~J1720.2+3536   & 17:20:12.0 &   +35:36:00 & 0.389 &  2008 Sep 3     &     69.6 & \\
CL~0024+16          & 00:26:35.7 &   +17:09:45 & 0.394 &  2007 Dec 18    &   $-$5.5 & \\
MACS~J0429.6$-$0253 & 04:29:41.1 & $-$02:53:33 & 0.400 &  2008 Sep 3     &     46.3 & \\
MACS~J0159.8$-$0849 & 01:59:00.0 & $-$08:49:00 & 0.405 &  2008 Sep 3     &     17.6 & \\
Abell 851           & 09:43:02.7 &   +46:58:37 & 0.405 &  2007 Dec 17    &     60.3 & \\
GHO~0303+1706       & 03:06:19.1 &   +17:18:49 & 0.423 &  2005 Feb 10    &  $-$60.0 & \\
MS~1621.5+2640      & 16:23:00.0 &   +26:33:00 & 0.428 &  2008 Jul 1     &     27.9 & \\
MACS~J1610.6+3810   & 16:21:24.8 &   +38:10:09 & 0.465 &  2008 Sep 3     &  $-$47.1 & \\
MACS~J0257.1$-$2325 & 02:57:09.1 & $-$23:26:06 & 0.505 &  2008 Sep 3     &     51.8 & \\
MS~0451.6$-$0306    & 04:54:10.8 & $-$03:00:57 & 0.539 &  2005 Feb 10    &  $-$39.1 & \\
MS~0451.6$-$0306    & 04:54:10.8 & $-$03:00:57 & 0.539 &  2007 Dec 17-18 &  $-$83.4 & \\
Bo\"otes 10.1       & 14:32:06.0 &   +34:16:47 & 0.544 &  2008 Sep 3     &     54.5 & \\
CL~0016+16          & 00:18:33.5 &   +16:25:15 & 0.545 &  2007 Dec 18    &    139.4 & very poor \\
MACS~J2129.4$-$0741 & 21:26:46.9 & $-$07:54:36 & 0.570 &  2008 Jul 1     &     75.0 & \\
MACS~J0025.4$-$1222 & 00:25:09.4 & $-$12:22:37 & 0.578 &  2008 Jul 1     &  $-$25.9 & \\
MACS~J0647.7+7015   & 06:47:50.5 &   +70:14:55 & 0.591 &  2009 Mar 2     &  $-$71.1 & mask 1 \\
MACS~J0647.7+7015   & 06:47:50.5 &   +70:14:55 & 0.591 &  2009 Mar 3     &  $-$95.5 & mask 2 \\
MACS~J0744.8+3927   & 07:44:52.5 &   +39:27:27 & 0.697 &  2009 Mar 2     &     91.7 & mask 1 \\
MACS~J0744.8+3927   & 07:44:52.5 &   +39:27:27 & 0.697 &  2009 Mar 3     &     97.6 & mask 2 \\
RCS~2318+0034       & 23:18:31.5 &   +00:34:18 & 0.756 &  2008 Sep 3     &  $-$26.1 & \\
Bo\"otes 10.8       & 14:32:06.0 &   +34:16:47 & 0.921 &  2008 Jul 1     &  $-$25.0 & \\
\enddata


\tablecomments{Cluster positions are in the J2000 coordinate system.}

\label{table.clusters}
\end{deluxetable}
\normalsize

\scriptsize
\begin{deluxetable}{ccccccl}
\tablecaption{Spectroscopic Results.}
\tablehead{
\colhead{Object ID} &
\colhead{RA} &
\colhead{Dec} &
\colhead{Redshift} &
\colhead{Quality} &
\colhead{Class} &
\colhead{Notes}}
\startdata
CL0024 gxy07  & 00:26:24.620 & +17:13:33.510 & 0.246 & A & S & [\ion{O}{2}],H$\alpha$ \\
CL0024 gxy26  & 00:26:25.420 & +17:13:22.812 & 0.399 & A & E & MgB \\
CL0024 g202p  & 00:26:26.029 & +17:11:16.559 & 0.392 & A & E+S & [\ion{O}{2}],CaHK \\
CL0024 g215p  & 00:26:27.117 & +17:12:26.082 & 0.398 & A & E & CaHK \\
CL0024 gxy21  & 00:26:27.709 & +17:13:49.797 & 0.395 & A & S & [\ion{O}{2}] \\
CL0024 g226p  & 00:26:27.956 & +17:11:37.879 & 0.394 & A & E & CaHK \\
CL0024 gxy06  & 00:26:28.381 & +17:12:46.550 & 0.397 & B & E & CaHK \\
CL0024 gxy04  & 00:26:31.029 & +17:12:09.808 &   0.0 & A & $\star$ & M-star \\
CL0024 gxy03  & 00:26:31.392 & +17:10:56.131 & 0.397 & A & E & CaHK \\
CL0024 g334p  & 00:26:34.210 & +17:10:09.844 & 0.387 & A & E & CaHK \\
CL0024 g338p  & 00:26:34.348 & +17:10:22.630 & 0.388 & A & E & CaHK \\
CL0024 gxy08  & 00:26:34.626 & +17:08:10.100 &   0.0 & A & $\star$ & M-star\\
CL0024 g353p  & 00:26:34.838 & +17:09:19.273 & 0.398 & A & E & CaHK \\
CL0024 g362p  & 00:26:35.018 & +17:09:39.309 & 0.399 & A & E & CaHK \\
CL0024 g363p  & 00:26:35.206 & +17:09:49.238 & 0.389 & A & E & CaHK \\
CL0024 g383p  & 00:26:36.140 & +17:08:27.616 & 0.402 & A & E & CaHK \\
CL0024 gxy02  & 00:26:36.250 & +17:10:00.925 & 0.390 & A & E & CaHK \\
CL0024 gxy09  & 00:26:37.313 & +17:07:50.647 & 0.392 & A & E & CaHK \\
CL0024 gxy10  & 00:26:38.418 & +17:07:31.566 & 0.392 & A & E & CaHK \\
CL0024 g458p  & 00:26:41.879 & +17:10:45.495 & 0.390 & A & E & CaHK \\
\enddata

\tablecomments{A sample from the full table is shown above; the
full table of 496 sources is included in the on-line version of the
paper.  Positions are in the J2000 coordinate system.  Typical
redshift uncertainties are $\pm 0.002$.  Only quality ``A'' (very
reliable) and ``B'' (reliable) redshifts are presented.  Spectroscopic
classes are E (elliptical, or early-type spectrum), S (spiral, or
late-type spectrum), AGN (quasar), and $\star$ (Galactic star).}

\label{table.spectra}
\end{deluxetable}
\normalsize

\clearpage
\end{document}